\documentclass[fleqn,usenatbib]{mnras}

\usepackage{newtxtext,newtxmath}

\usepackage[T1]{fontenc}
\usepackage{ae,aecompl}

\usepackage{graphicx}	
\usepackage{amsmath}	
\usepackage{amssymb}	

\title[Stellar noise in the ARIEL space mission.]{Stellar pulsation and granulation as noise sources in exoplanet transit spectroscopy in the ARIEL space mission.}

\author[S. Sarkar et al.]{
Subhajit Sarkar,$^{1}$\thanks{E-mail: subhajit.sarkar@astro.cf.ac.uk (SS)}
Ioannis Argyriou$^{2}$, Bart Vandenbussche$^{2}$, 
\newauthor{
Andreas Papageorgiou$^{1}$ and Enzo Pascale$^{1,3}$}\\
$^{1}$School of Physics and Astronomy, Cardiff University, The Parade, Cardiff, CF24 3AA, UK\\
$^{2}$Institute of Astronomy, KU Leuven, Celestijnenlaan 200D Box 2401, 3001 Leuven, Belgium\\
$^{3}$Department of Physics, La Sapienza University  of Rome, Piazzale Aldo Moro 2, 00185 Rome, Italy}

\date{Accepted XXX. Received YYY; in original form ZZZ}

\pubyear{2018}

\begin{document}
\label{firstpage}
\pagerange{\pageref{firstpage}--\pageref{lastpage}}
\maketitle

\begin{abstract}
Stellar variability from pulsations and granulation presents a source of correlated noise that can impact the accuracy and precision of multi-band photometric transit observations of exoplanets.  This can potentially cause biased measurements in the transmission or emission spectrum or underestimation of the final error bars on the spectrum.  ARIEL is a future space telescope and instrument designed to perform a transit spectroscopic survey of a large sample of exoplanets. In this paper we perform simulations to assess the impact of stellar variability arising from pulsations and granulation on ARIEL observations of GJ 1214b and HD 209458b.  We take into account the correlated nature of stellar noise, quantify it, and compare it to photon noise.  In the range 1.95-7.8 \textmu m, stellar pulsation and granulation noise has insignificant impact compared to photon noise for both targets.  In the visual range the contribution increases significantly but remains small in absolute terms and will have minimal impact on the transmission spectra of the targets studied.  The impact of pulsation and granulation will be greatest for planets with low scale height atmospheres and long transit times around bright stars.
\end{abstract}




\section{Introduction}

Stellar variability as source of astrophysical noise can impact the precision and accuracy of exoplanet multi-band photometric transit observations.  ARIEL (Atmospheric Remote-sensing Exoplanet Large-survey) is the ESA Cosmic Vision M4 mission which aims to perform the first space-based spectroscopic large sample survey of exoplanets \citep{Tinetti2016}.  In this paper we attempt to predict the impact of stellar variability resulting from pulsations and granulation on ARIEL observations specifically, and present a methodology that can be applied to other instruments and a range of possible targets.
  
Convection in the outer layer of stars induces pulsations and causes granulation. Pulsations arise from propagation of pressure waves causing expansion and contraction of the outer layer. They can be represented as a spectrum of oscillatory 'p-modes' with time scales of the order of 3-15 min in solar-like oscillators \citep{Bedding2003}. Granules form from convection currents of plasma, with cells having size of the order of 10$^{6}$ m \citep{Micela2015} and lifetimes of 10 to 22 min \citep{Zirin1992}. For the radial velocity method, previous simulations have shown a non-negligible  impact of noise from stellar pulsations and granulation with mitigation strategies suggested, e.g \cite{Dumusque2011}. The impact on transit observations has been mostly focused on photometric transits and pulsation noise in highly variable stars such  as the  $\delta$ Scuti star, WASP-33, with its highly inflated hot Jupiter companion. By analysis of the out-of-transit light curve, a model of pulsation with multiple p-modes can be constructed and used to decorrelate the data. 
 \cite{vonEssen2014}, looking at visual range photometric transits, found that accounting for pulsations did not result in any significant change in the transit parameters but improved the precision on the radius ratio, $R_p/R_s$, by 25\%. The effects of granulation noise on transit multi-band photometry was assessed by \cite{Chiavassa2017} using a radiative hydrodynamical model of stellar granulations. They quantified the 1$\sigma$ uncertainty on the planet radius arising from granulation. For example, for a hot Jupiter-Sun system this was 0.05-0.36\% depending on the wavelength. However the study did not include pulsations or compare the noise to the photon noise limit.
\\ \indent Given the paucity of studies regarding the impact of stellar pulsations and granulation on exoplanet transit spectroscopy, we adopt the novel approach of combining a model of stellar variability that includes both pulsation and granulation with an end-to-end simulator of transit spectroscopy observations to produce realistic simulated observational timelines. Unlike previous studies the noise from stellar pulsation and granulation is assessed in the context of the observational photon noise limit.

\vspace{-0.5em} 
\section{Models}

 \begin{figure}
\vspace{1.0em} 
	\includegraphics[trim={0cm 0cm 0cm 2.5cm}, clip, width=1.0\columnwidth]{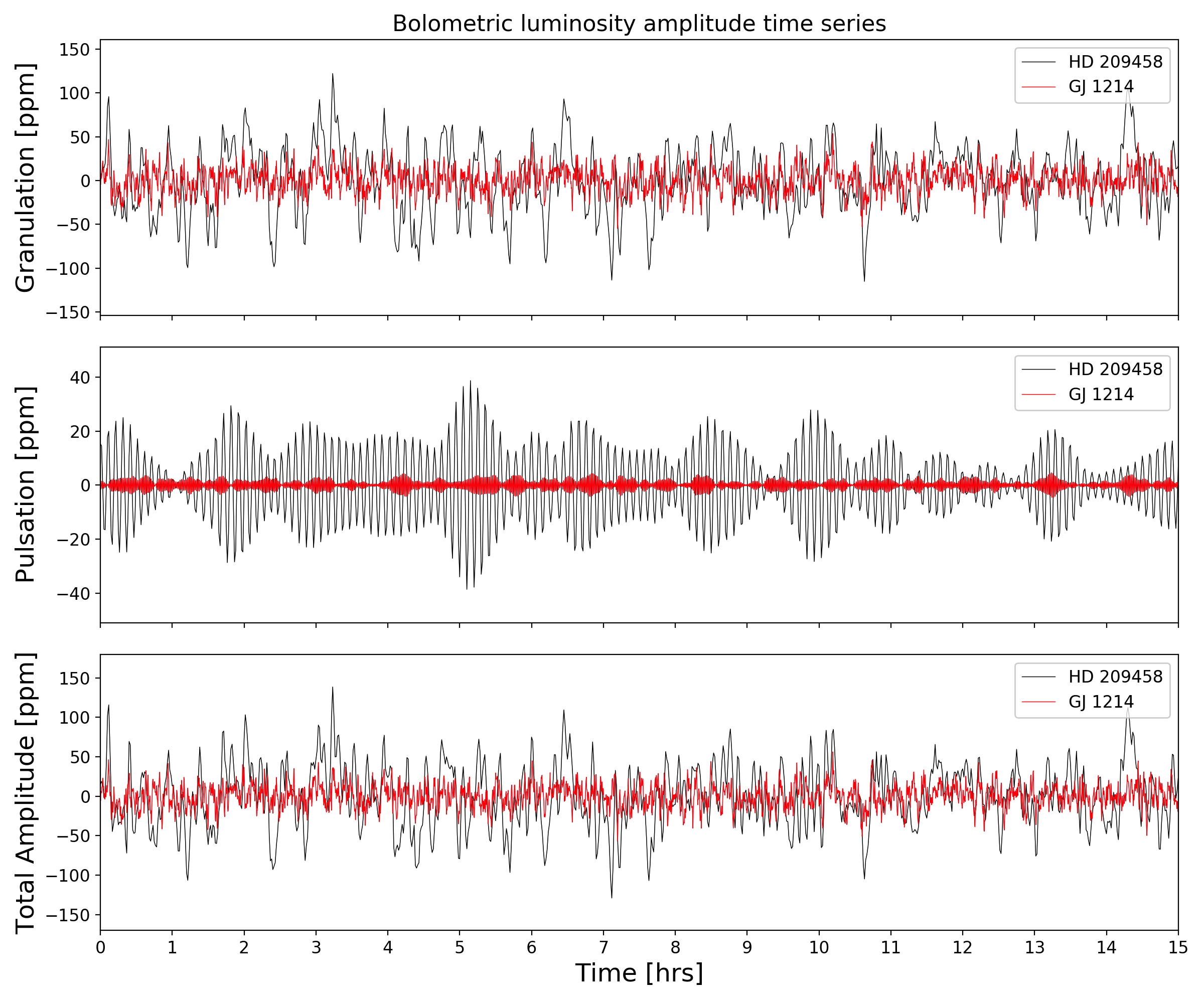}
\vspace{-1.0em} 
    \caption{Stellar variability model example time series for GJ 1214 (M4.5V) and HD 209458 (G0V).  One realization of 15 hours is shown for each star (60 s cadence for HD 209458 and 15 s cadence for GJ 1214; the reason for these different cadences is given in the text).  Displayed are the variations to the bolometric luminosity with time due to granulation (top), pulsation (middle) and their combined effects (bottom).}
    \label{fig:yannis}
\end{figure}

\begin{figure}
 \begin{center}
 	\includegraphics[width=1.0\columnwidth]{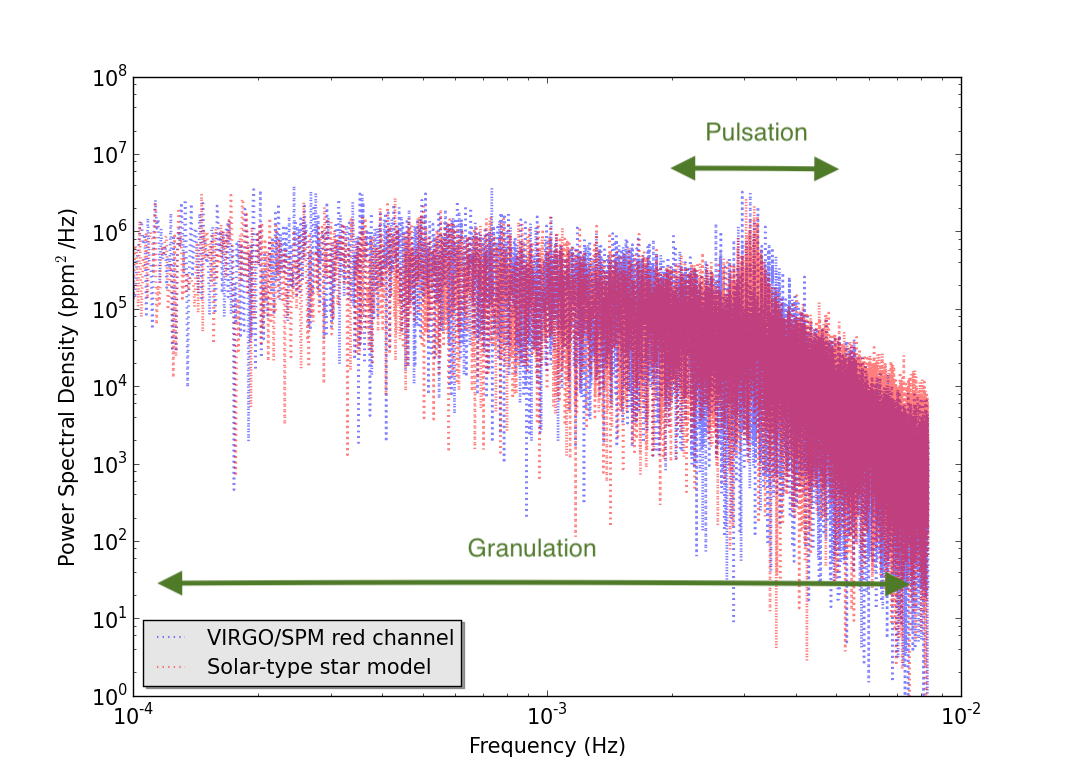}
 \end{center}
\vspace{-0.3em}  
    \caption{Comparison between solar-type star model and VIRGO/SPM red channel (0.86 \textmu m) power spectral density (PSD) profiles.  For the model, 20 realizations of 15 hours and 60 s cadence each were combined to give a total SED timeline of 300 hours.  From this the 0.86 \textmu m (bandwidth 0.01 \textmu m) timeline was selected and transformed to obtain the PSD shown. For comparison, a 300 hour and 60 s cadence sequence was chosen arbitrarily from publicly available VIRGO/SPM red channel data from the year 2002, with the displayed PSD constructed from this.
Pulsation modes correspond to the region of the high frequency 'bump', whereas granulation contributes over the whole frequency range.}
    \label{fig:PSD}
\end{figure}

\subsection{Stellar variability model}

The effect of pulsations on a star's bolometric luminosity is modelled as follows. Pulsations are modeled using pulsation mode frequencies and decay times from BiSON solar data \citep{Broomhall2009}. For a given star, the pulsation mode frequencies are rescaled using the scaling relations from \cite{Kjeldsen1995}, and $\nu_{max}$ (the frequency of maximum pulsation)
and $\Delta \nu_o$ (the frequency spacing between pulsation modes) are computed. We use $3.14 \times 10^3$ \textmu{}Hz for the solar value of $\nu_{max}$ \citep{Kallinger2014}.  The pulsation mode rms amplitudes are assumed to have a Gaussian distribution, with the peak of the Gaussian, $A_o$, calculated using the scaling relation given by equation 26 of \cite{Corsaro2013}, and centred at frequency $\nu_{max}$. The standard deviation of the distribution, $s_o$, is set to $1.5 \times \Delta \nu_o$.  Each pulsation mode is simulated as timeline of stochastically excited damped oscillations, based on the model described in \cite{DeRidder2006}. The timelines are then co-added. A time series of bolometric luminosity variations due to pulsations is thus generated.  

Granulation signals are first simulated in Fourier space with a power spectral density (PSD) profile, $P_g(\nu)$.  We follow the formulation of \cite{Kallinger2014} where $P_g(\nu)$ is the sum of 2 super-Lorentzian functions. The super-Lorentzian profile is a modified version of the Harvey model, formerly used to study solar granulation signals \citep{Harvey1985}. Each of $j$ super-Lorentzian functions ($j=1,2$) has a characteristic frequency, $b_j$, and an rms amplitude, $a_j$.  We use the empirical scaling relations from \cite{Kallinger2014} to compute the values of $a_1$, $a_2$, $b_1$ and $b_2$ for each star. A bolometric correction, $C_{bol}$, is applied to $a_1$ and $a_2$, where $C_{bol} = (T_{\star}/5934)^{0.8}$ and $T_{\star}$ is the stellar temperature \citep{Kallinger2014}.  The relations used are: $a1 = a2 = C_{bol}(3710 \cdot {\nu_{max}}^{-0.613}{M_{\star}}^{-0.26}$) ppm, $b_1 = 0.317 \cdot {\nu_{max}}^{0.970}$ \textmu{}Hz, and  $b_2 = 0.948\cdot{\nu_{max}}^{0.992}$ \textmu{}Hz, where $M_{\star}$ is the mass of the star in solar masses, and $\nu_{max}$ is the maximum pulsation frequency in \textmu{}Hz. $P_g(\nu)$ is expressed in the second RHS term in Eq. \ref{eq:1}, where the factor $\xi$ is a normalizing constant set to $2 \sqrt{2}/ \pi$.  For each realization, a random timeline of bolometric luminosity variations is generated as follows. The rms amplitude of a sinusoidal component at any frequency $\nu$ is given by $A_g(\nu) = \sqrt{p_g(\nu)} = \sqrt {P_g(\nu) \Delta\nu}$, where $p_g(\nu)$ is the power and $\Delta\nu$ is the frequency spacing in the PSD. In reality, a noisy timeline occurs since each frequency component will have variations in its rms amplitude and phase over time. The average power over time will remain $p_g(\nu) = {A_g(\nu)}^2$.  
We assume that the random variation in rms amplitude follows a normal distribution, ${A_g}'(\nu) \sim N(0, {A_g(\nu)}^2)$, i.e. a mean of 0 and a standard deviation of $A_g(\nu)$, and that the random variation in phase angle follows a uniform distribution $\sim U(0,2\pi)$.  The distribution ${{A_g}'(\nu)}^2$ (the squared values of the above normal distribution) has a mean value that equals the expected power $p_g(\nu)$.  As transformation of time varying amplitude and phase spectra is challenging, we approximate this effect by randomising the spectra once at the start of each realization. This is performed as follows. For each value of $A_g(\nu)$, a random complex number is generated, ${A_g}''(\nu)$, such that ${A_g}''(\nu) = A_g(\nu)(x + iy)$, where $x$ and $y$ are independent normally distributed random variables with mean 0 and standard deviation $1/\sqrt{2}$.  The modulus of ${A_g}''(\nu)$ is thus a normally distributed random variable with mean 0 and standard deviation of $A_g(\nu)$, and the phase of ${A_g}''(\nu)$ is a uniformly distributed random variable $\sim U(0,2\pi)$.  The inverse real Fourier transform is then performed on ${A_g}''(\nu)/ \sqrt{2}$, and normalised by multiplying by $2(N_{f}-1)$, where $N_{f}$ is the number of samples in the power spectrum, to generate a random timeline. We confirm the validity of this method by verifying that the variance of the randomly generated noisy timeline approximately equals the integral of $P_g(\nu)$ (Parseval's theorem). The model thus outputs a time series of bolometric luminosity variations due to granulation. 

The granulation and pulsation bolometric luminosity variations are then summed (Figure 1). For the overall model, the average power spectral density at any frequency $\nu$ can be summarized as follows:
\vspace{0.5em}
\begin{equation}
\label{eq:1}
\begin{split}
P(\nu) = \sum_{i=1}^{N_o} \zeta_i(\nu) &\left[A_o \exp \left( \frac{- [ \nu_i - \nu_{max} ] ^2 }{ 2 {s_o}^2} \right)\right]^2  \\
 & \quad \quad \quad \quad \quad \quad \quad + \sum_{j=1}^{2} \left[\frac {\xi {a_j}^2/b_j} { 1+ (\nu /b_j)^4} \right]
\end{split}
\end{equation}
\vspace{-0.2em}

\noindent where the first term on the RHS is the pulsation component, and the second term is the granulation component, $P_g(\nu)$, described above. The pulsation component is the result of summing the PSD of $i$ pulsation modes ($i = 1,2,...N_o$, where $N_o$ is the total number of modes). The rms amplitude of the $i$th mode is given by $A_o \exp(-[\nu_i - \nu_{max}]^2/2{s_o}^2)$, where $\nu_i$ is the oscillation frequency. The function $\zeta_i(\nu)$ accounts for the distribution of power from each pulsation mode.  The PSD of such a stochastic damped oscillation mode can be described by a Lorentzian function \citep{DeRidder2006} so that:
$\zeta_i(\nu)  = \Gamma / \pi [ 4(\nu-\nu_i)^2 + \Gamma^2]$. $\Gamma$ is the half width at half maximum (HWHM) of the Lorentzian profile where $\Gamma = \eta/\pi$, and $\eta$ is the damping rate.

A time-dependent variation in effective temperature is then computed from the bolometric luminosity amplitude variation as follows. The local gradient of luminosity change against temperature change is found by taking  a fine grid of PHOENIX NextGen stellar spectra \citep{Hauschildt1999} sharing the same metallicity and surface gravity, and selecting a pair of models with temperatures just above and just below the nominal effective temperature of the star.  The associated luminosities are then calculated by integrating each spectrum and multiplying by the stellar surface area, thus obtaining the local gradient. It is assumed that within the time scale of a transit the variation in stellar radius is negligible. By adding the variation in effective temperature to the nominal effective temperature of the star of interest, an effective temperature time series is created. Subsequently, a linear interpolation between the above stellar models is carried out, at the effective temperature value associated with each time datum. This produces a time series of spectral energy distributions (SEDs).

A comparison between the model (using solar parameters) and data from the Solar and Heliospheric Observatory (SOHO) VIRGO/SPM instrument \citep{Frohlich1995, Frohlich1997, Jiminez1999} is made in Figure 2, which shows the power spectral density (PSD) profiles for each. Some discrepancies may occur from the scaling relations used which are calibrated from large samples of stars, and the limits of the modelling process, however there is in general a good match between the model and the real data.

For this study, the above stellar variability model was used to simulate SED time series for GJ 1214 (an M4.5V star)  and HD 209458 (a G0V star).  For HD 209458, we used a 60 s timebase in order to incorporate the pulsation component which peaks at $\nu_{max} = 2.7 \times 10^{3}$ \textmu{}Hz.  For GJ 1214, we used a 15 s timebase as the pulsation contribution peaks at $\nu_{max} = 1.4 \times 10^{4}$ \textmu{}Hz.   Both the pulsation peak and the granulation rms amplitude are inversely related to $\nu_{max}$ \citep{Corsaro2013, Kallinger2014}.

\vspace{-0.5em} 
\subsection{ExoSim}

ExoSim is a generic end-to-end simulator of transit spectroscopy observations, that can be applied to different instruments and targets \citep{Sarkar2016}. ExoSim models the time domain in small steps, and thus can capture the effects of time-correlated noise sources and systematics. ExoSim includes models of both the astrophysical scene, including the star and planet, and the entire optical chain including telescope, instrument and detector.  

The inputs to ExoSim are defined in a single \textit{input configuration file} (ICF) that contains the observational and instrumental parameters for the simulation.  Both spectroscopic and photometric channels can be simulated. The target exosystem is chosen within the ICF and the relevant stellar and planetary parameters are then automatically selected from the Open Exoplanet Catalogue \citep{Rein2012} database.  It also selects the best matching PHOENIX stellar model for the host star. The length of the observation, exposure time, number of non-destructive reads, and other observational parameters can be set in the ICF. The telescope, instrument channels and detectors are also defined in the ICF and link to reference files for specific characteristics such as the wavelength solution in each channel and transmissions and emissivities of each optical element.

ExoSim takes the baseline stellar signal and modulates this through its astrophysical and instrument models. It can apply multiple noise sources, systematics, and the effects of time-dependent processes such as stellar pulsation and granulation.  ExoSim outputs simulated time series images akin to a real observation which then require a data reduction pipeline.  It can generate either full transit or out-of-transit simulations.  In this case we elected to run ExoSim in out-of-transit (OOT) mode, as this was adequate for the comparison between photon noise and stellar variability noise.

\vspace{-0.9em}
\section{Simulations and data reduction}

The main parameters of the ARIEL model used are given in Table \ref{table:ariel}. This represents the configuration at the end of Phase A with further details given in \cite{Sarkar2017}. The main ARIEL spectrometer (AIRS) covers the wavelength range 1.95-7.8 \textmu{}m in two channels: Ch0 (1.95-3.9 \textmu{}m with spectral resolving power R $\geq$ 100) and Ch1 (3.9-7.8 \textmu{}m at R $\geq$ 30) and will conduct the bulk of the exoplanet science obtaining transmission and emission spectra.  A low resolution near-infrared spectrometer (NIRSpec) and 3 visual photometric channels are also included (Table \ref{table:ariel}) that can constrain the Rayleigh scattering slope and monitor for stellar variations. The latter are, in addition, used in the fine guidance system for the spacecraft attitude control system.

\begin{table}
\begin{center}
\caption{ARIEL configuration used in ExoSim. The main IR spectrometer (AIRS) is divided into Ch0 and Ch1. NIRSpec is a NIR low-resolution spectrometer. VisPhot, FGS 1, and FGS 2 are photometric channels. $\lambda$ is wavelength coverage, f-number is image space focal ratio, $\upsilon$ is mid-band transmission, QE is detector quantum efficiency, $\delta$ is pixel length and $\phi$ is plate scale.}
\label{table:ariel}

\setlength{\tabcolsep}{5pt}
\begin{tabular}{lcccccc} 
\hline
Channel&Vis-&FGS&FGS&NIR-&AIRS&AIRS\\
&Phot&1&2&Spec&Ch0&Ch1 \\
\hline
$\lambda$ (\textmu m)&
0.5&
0.9&
1.1&
1.25-&
1.95-&
3.9-\\
&
&
&
&
1.9&
3.9&
7.8\\
Prism R&
N/A&
N/A&
N/A&
40-&
100-&
30-\\
&
&
&
&
60&
160&
70\\
Binned R&
N/A&
N/A&
N/A&
20&
100&
30\\
f-number&
39.5&
24.6&
31.3&
19.3&
13.2&
6.36\\
$\upsilon$&
0.53&
0.60&
0.60&
0.65&
0.45&
0.46\\
QE&
0.55&
0.55&
0.55&
0.55&
0.55&
0.55\\
$\delta$ (\textmu m)&
18 &
18 &
18 &
18 &
15 &
15 \\
$\phi$ $(^\circ \times 10^{-5}/\delta)$&
2.86 &
4.58 &
3.60 &
4.85 &
6.11 &
1.23 \\
\hline
\end{tabular}
\end{center}
\end{table}

SED timelines from the stellar variability model were obtained for the stars HD209458 and GJ1214.  These were used for modelling the flux variations of the host stars of the exoplanets HD 209458b (a 'hot Jupiter') and GJ 1214b (a 'Super-Earth') respectively.  The model SED timelines were rebinned to the wavelength solution of each ARIEL channel for the ExoSim simulation, where they were used to proportionally modulate the wavelength-dependent flux of the star with time.  Although ExoSim can model a variety of different noise sources, in this study we simulate only stellar variability noise (from combined granulation and pulsation), and the source photon noise, each in isolation.  Since read out noise is not under evaluation and we use 100\% exposure duty cycle, the time chosen for each exposure in ExoSim does not affect the final noise when binned over several exposures, and is thus set at 60 s for all channels and sources.

For each stellar target, ExoSim simulated an OOT observation of 15 hours (the duration of the stellar variability timelines used) consisting of 60 s exposures.  This was obtained initially for photon noise from the source only, and then for stellar variability noise.  20 realizations were simulated for each case.  

For each channel the time series of spectral or photometric images were processed using a standardised ExoSim pipeline described in detail in \cite{Sarkar2017}, consisting of the following steps: 1) flat fielding  2) background subtraction 
3) correlated double sampling  4) jitter decorrelation (omitted for non-jitter sources as here) 5) aperture masking  6) binning into spectral-resolution-element-sized bins (for spectroscopic channels) and 7) extraction of 1-D spectra per exposure for spectroscopic channels, or the photometric aperture count for photometric channels.  The apertures used here spanned the width of the Airy disc or its equivalent energy.
The output of the pipeline gives the photoelectron count in a spectral bin (or photometric aperture) per exposure in the time series.

\vspace{-1.0em} 
\section{Results}

\begin{figure}
	\includegraphics[trim={0 0 0 0}, clip,width=\columnwidth]{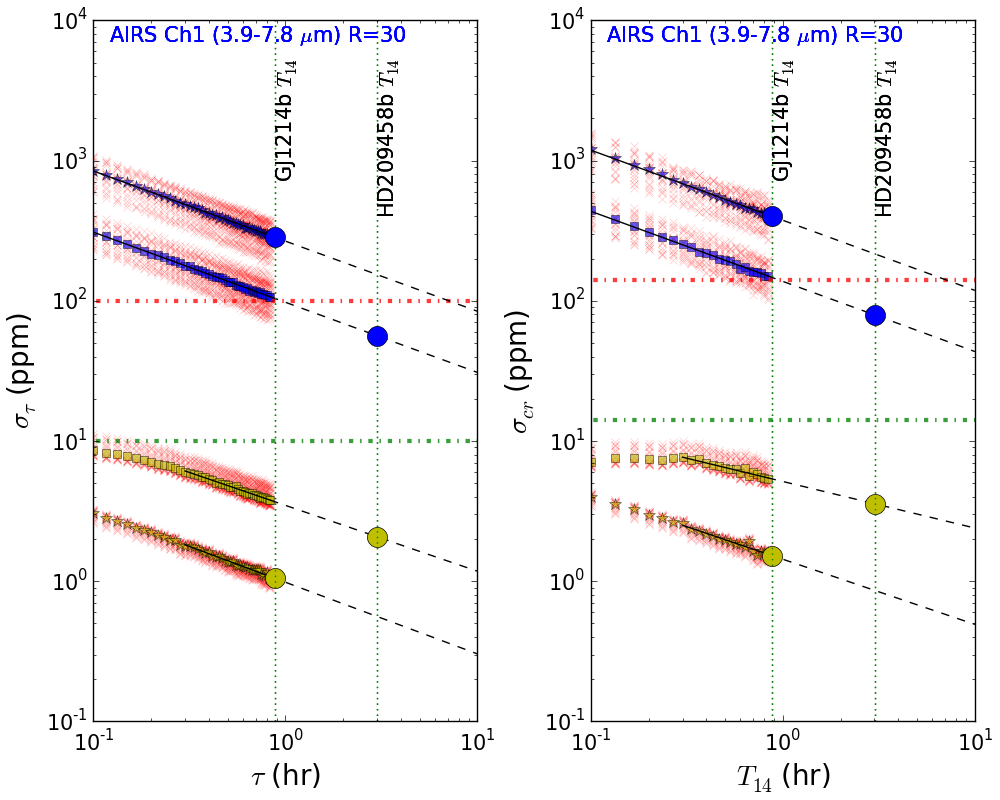}
	\includegraphics[width=\columnwidth]{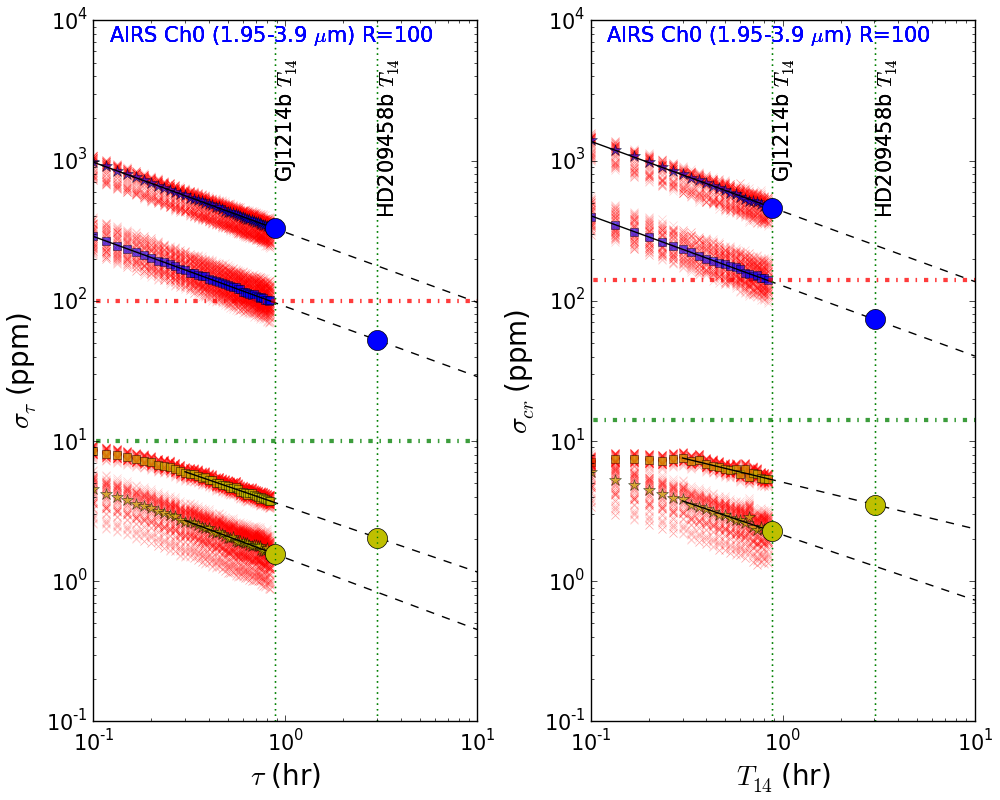}
	\includegraphics[width=\columnwidth]{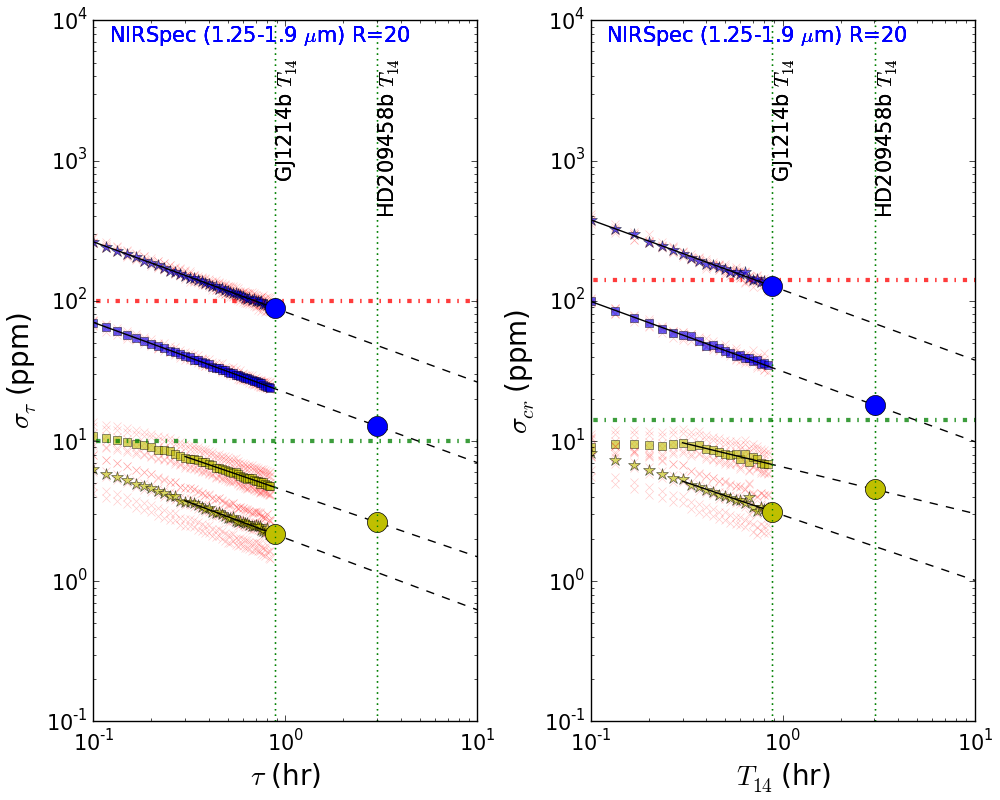}
\vspace{-1.5em}
    \caption{ARIEL spectroscopic channels. Left: $\sigma_{\tau}$ vs $\tau$.  Right: $\sigma_{cr}$ vs $T_{14}$.  Red crosses: individual spectral bins (number of bins highest in Ch0 and lowest in NIRSpec, reflecting the binned R power and wavelength range in each channel). Blue stars/squares: photon noise. Yellow stars/squares: pulsation \& granulation noise. Stars: GJ 1214. Squares: HD 209458. Black solid/dashed lines: power law fits/extrapolations. Large dots: noise expected at the time scale of the planet transit. Upper red \& lower green dot-dashed line show ARIEL 'noise floor' requirement and goal respectively ($\times \sqrt{2}$ for $\sigma_{cr}$).}
    \label{fig:comp1}
\end{figure}

\begin{figure}

	\includegraphics[width=\columnwidth]{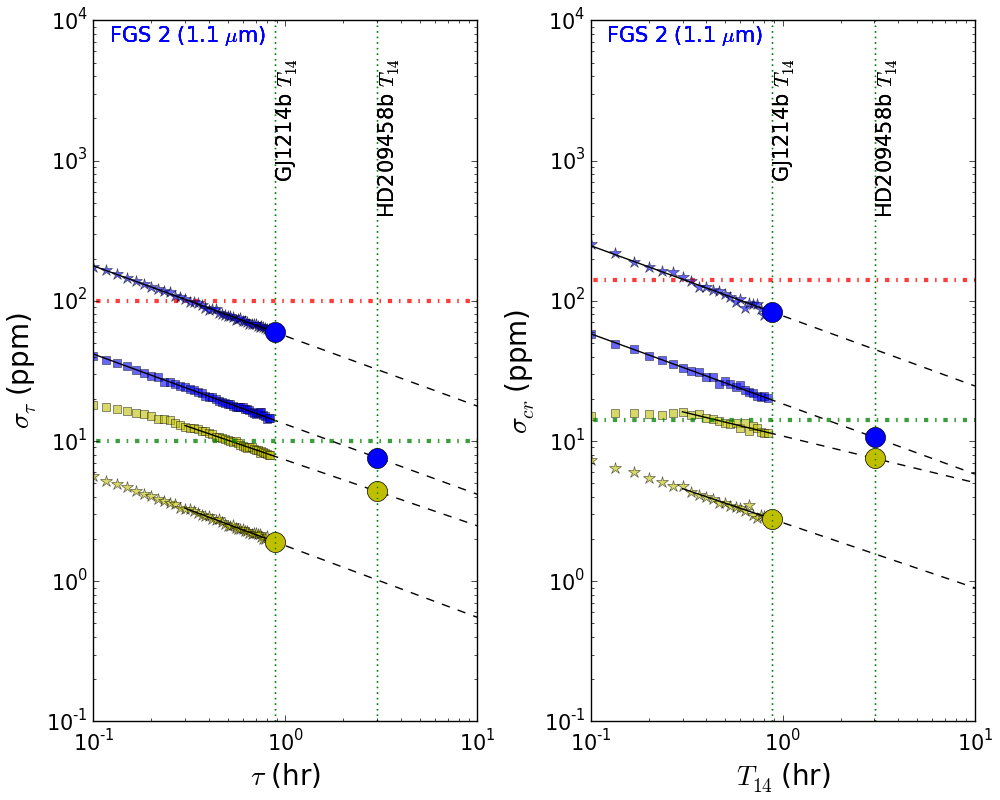}
	\includegraphics[width=\columnwidth]{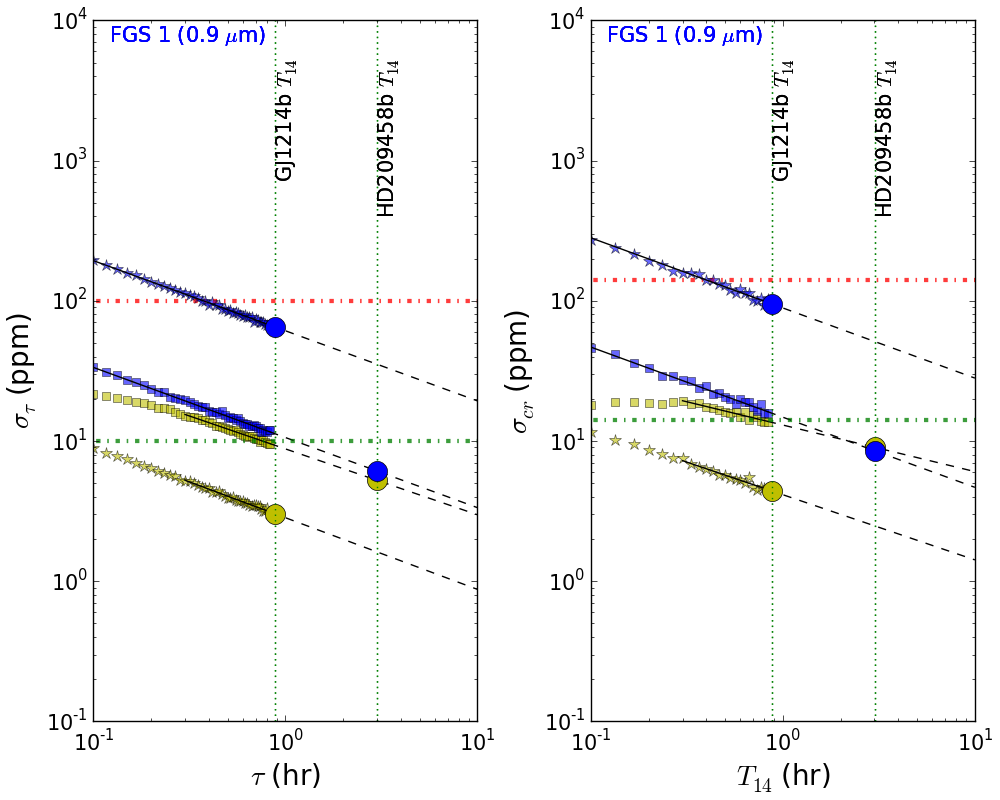}
	\includegraphics[width=\columnwidth]{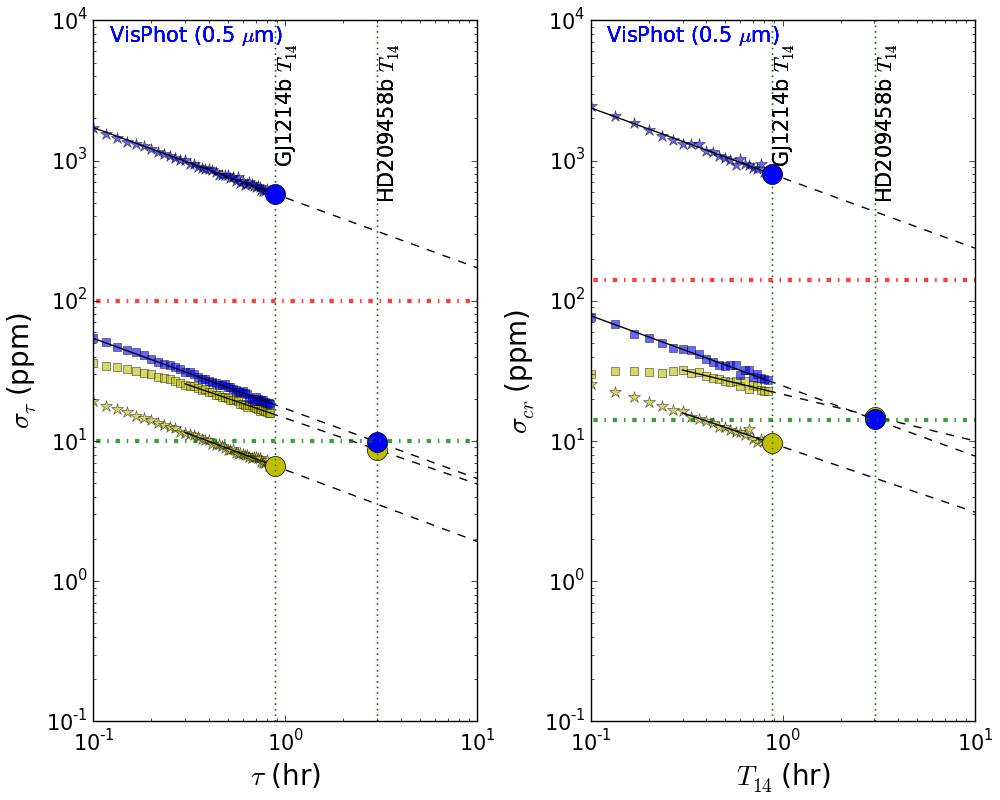}
\vspace{-1.5em}
    \caption{ARIEL photometric channels.  Left: $\sigma_{\tau}$ vs $\tau$.  Right: $\sigma_{cr}$ vs $T_{14}$. Blue stars/squares: photon noise. Yellow stars/squares: pulsation  \& granulation noise. Stars: GJ 1214b. Squares: HD 209458. Black solid/dashed lines: power law fits/extrapolations. Large dots: noise expected at the time scale of the planet transit. Upper red dot-dashed line and lower green dot-dashed line show ARIEL 'noise floor' requirement and goal respectively ($\times \sqrt{2}$ for $\sigma_{cr}$).}
    \label{fig:comp2}
    
\end{figure}

For each stellar target and noise source in each channel we obtain results for the fractional uncertainty on the stellar flux, $\sigma_{\tau}$, at different binned integration periods ($\tau$) (Figures \ref{fig:comp1} and \ref{fig:comp2}, left panels), which can be considered the 'photometric precision'.  We also obtain for each noise source in each channel the uncertainty on the contrast ratio, $\sigma_{cr}$, at different possible values of the transit duration, $T_{14}$ (Figures \ref{fig:comp1} and \ref{fig:comp2}, right panels). $\sigma_{cr}$ represents the 1$\sigma$ error on the measured contrast ratio (or fractional transit depth), $(R_p/R_s)^2$, where $R_p$ and $R_s$ are the planet and star radius respectively. The spectrum of contrast ratios with wavelength, $(R_p/R_s)^2(\lambda)$, constitutes the planet transmission spectrum.

$\sigma_{\tau}$ is found as follows. For each spectral bin or photometric channel, the 20 simulated exposure timelines from ExoSim (each of total length 15 hours duration and 60 s cadence) are combined to give a total timeline of 300 hours (1080000 s).  The simulation could not be run for 300 hours in one realization due to computational restrictions, but combining the timelines gives more overall data points to allow for robust assessment of the standard deviation and time-dependent behaviour.  The combined timeline is then divided into consecutive bins, each of total duration $\tau$.   Each bin thus consists of several 60 s exposures.
Any bins that cross between two adjacent realizations are rejected to ensure only contiguous exposures from the same realization fall in each individual bin.  
  $\tau$ is varied from a lower limit of 360 s (6 exposures per bin) upto an upper limit of 3000 s (50 exposures per bin).  The lower limit was set 
so that the trend of  $\sigma_{\tau}$ with $\tau$ becomes apparent and we can determine the start of the stable trend at longer durations of $\tau$, permitting extrapolation of  $\sigma_{\tau}$ to timescales of the transit duration.
The upper limit was set by the need to obtain an adequate number of samples to measure the standard deviation. 
For each value of $\tau$, $\sigma_{\tau}$ is obtained as follows. In each bin, the mean value of the signal per exposure, $S$, is obtained, and then the standard deviation of $S$, $\sigma_s$, is calculated.  The fractional noise, $\sigma_{\tau}$, is then obtained by dividing $\sigma_s$ by the mean of $S$. Thus we obtain values of $\sigma_{\tau}$ versus $\tau$ for each spectral bin in each spectroscopic channel, and for each photometric channel (Figures \ref{fig:comp1} and \ref{fig:comp2}, left panels).  The median results are then obtained for each spectroscopic channel.
To these medians, and to the single results in each photometric channel, linear fits are performed in log-log space from which a power law relationship of $ \sigma_{\tau}$ with $\tau$ at long durations of $\tau$ can be estimated for each channel.
For the uncorrelated photon noise, $\sigma_{\tau,pn}$  ($pn$ = 'photon noise'),  the power law exponent $\alpha$ should be -0.5. This follows from the fact that for uncorrelated noise the standard error on the mean falls with the square root of the size of the sample.  Since $\tau$ is proportional to sample size (number of exposures per bin), we should expect $\sigma_{\tau,pn}$ to vary inversely with $\sqrt{\tau}$, hence giving a value for $\alpha$ of -0.5.  Thus for the photon noise, the results are fitted with a power law model where $\alpha$ was fixed to -0.5. 
For the correlated stellar pulsation and granulation noise, $\sigma_{\tau,sn}$ ($sn$ = 'stellar noise'), this relationship however may not hold and therefore $\alpha$ is allowed to vary as a free parameter.   
We see in Figures \ref{fig:comp1} and \ref{fig:comp2} (left panels), that for both stars, a stable trend at long durations appears established at points where $\tau \geq$ 1080 s and therefore the linear fits are performed only to these points.  The values of $\alpha$ for stellar noise are -0.51 ($\pm$ 0.01) for  GJ 1214 and -0.47 ($\pm$ 0.01)  for HD 209458, and are consistent across all channels. 
We extrapolate the power law models out to maximum $\tau$ of 10 hours, and use these to find $\sigma_{\tau}$ for each noise type when $\tau=T_{14}$ for the two planets (Figures \ref{fig:comp1} and \ref{fig:comp2}, left panels). $T_{14}$ is 3161 s (0.88 hours) and 10819 s (3.01 hours) for GJ 1214b and HD 209458b respectively.

The above method determines how the photon noise and the stellar variability noise integrate down with time if simple binning is considered, and gives a measure of photometric precision.  The contrast ratio, $(R_p/R_s)^2$, however depends, to a first order, on measuring the difference between in- and out-of-transit signal means.   For correlated noise, additional correlation may exist between the in- and out-of-transit data, meaning that the uncertainty on the contrast ratio,  $\sigma_{cr}$, may not follow the same time-dependency or power law relationship as $\sigma_{\tau}$.  For uncorrelated noise, e.g. photon noise, since there is no correlation between in- and out-of-transit data, we would expect  $\sigma_{cr}$ to have the same power law relationship as $\sigma_{\tau}$, i.e. $\alpha$ = -0.5.  Simply considered the transit depth, $\rho$, is the difference of two mean values: the in-transit mean ($\mu_1$) and the out-of-transit mean ($\mu_2$).  The contrast ratio is the fractional transit depth, $\rho/\mu_2$, with uncertainty $\sigma_{cr}$.  If $\tau = T_{14}$, then for uncorrelated photon noise, $\sigma_{cr}$ can be approximated by $\sqrt{2} \sigma_{\tau}$, assuming an equal amount of time observed in- and out-of-transit. However this relationship may not hold for correlated noise. 

We therefore estimate $\sigma_{cr}$ for a single transit observation using the following approach, which assumes an equal amount of time observed in- and out-of-transit, and the out-of-transit portion equally divided pre- and post-transit.  The combined ExoSim timelines (of total duration 300 hours), for each stellar target and noise type in each spectral bin or photometric channel, are again divided into consecutive bins, but this time each of duration 2 $\times$ $T_{14}$.  $T_{14}$ is varied from 360 s upto 
3000 s for the same reasons as given above.  Again, any bins that cross between two adjacent realizations are rejected to ensure only contiguous exposures from the same realization fall in each individual bin.   Each bin is then sub-divided into three consecutive sections.  The first section, $x_1$, consists of exposures amounting to total duration $T_{14}/2$.  The middle section, $x_2$, consists of exposures amounting to total duration $T_{14}$.  The final section, $x_3$, consists of exposures amounting to total duration $T_{14}/2$.  For each bin, $\mu_1$ is the mean signal in the exposures making up the middle section, $x_2$, and represents the 'in-transit' mean.  $\mu_2$ is the mean signal in the exposures making up the two outer sections, $x_1$ and $x_3$, and represents the 'out-of-transit' mean.  For each bin, the difference $\rho = \mu_2 - \mu_1$, is obtained.  $\rho$ would be zero in the absence of noise.  The standard deviation of $\rho$, $\sigma_\rho$, is then found over all the bins comprising the timeline.  This is taken to be an estimate of the noise on the transit depth, since the same noise would be obtained if a fixed transit depth was present.  Obtaining $\sigma_\rho$ in this way maintains the effect of any correlation between the in- and out-of-transit data.  $\sigma_{cr}$ is then obtained by dividing $\sigma_\rho$ by $S$, the mean signal from all exposures in the timeline.  This is repeated for the range of values of $T_{14}$, so that values of $\sigma_{cr}$ versus $T_{14}$ are found for each spectral bin in each spectroscopic channel, and for each photometric channel (Figures \ref{fig:comp1} and \ref{fig:comp2}, right planels). 
As before, we find the median results for each spectroscopic channel. To these medians and to the single results in each photometric channel, linear fits are performed in log-log space from which the power law relationship of $\sigma_{cr}$ with $T_{14}$ at long durations of $T_{14}$ can be estimated for each channel.  For photon noise, $\sigma_{cr,pn}$, the fits are again obtained with $\alpha$ fixed to -0.5 for the reasons explained above.  As predicted $\sigma_{cr,pn}$ is $\sim \sqrt{2} \sigma_{\tau,pn}$.  For stellar variability noise, $\sigma_{cr,sn}$, $\alpha$ is allowed to vary, and linear fits performed to points at $T_{14} \geq$ 1080 s for the same reasons as before.
The fits return values for $\alpha$ of -0.47 ($\pm$ 0.03)  for  GJ 1214 and -0.33 ($\pm$ 0.02) for  HD 209458, consistent across all channels.  These results indicate that, particularly in HD 209458, $\sigma_{cr,sn}$ integrates down with $T_{14}$ at a slower rate than  $\sigma_{cr,pn}$.  The ratio of $\sigma_{cr,sn}$ to $\sigma_{cr,pn}$ therefore increases with $T_{14}$ for HD 209458.   In GJ 1214, at long integration periods, $\sigma_{cr,sn}$ has time-dependent behaviour closer to that of uncorrelated 'white' noise.  The linear fits are extrapolated out to a maximum $T_{14}$ of 10 hours, and from these, $\sigma_{cr}$ is found for each noise type at $T_{14}$ for the two planets, GJ 1214b and HD 209458b (Figures \ref{fig:comp1} and \ref{fig:comp2}, right panels).  

To assess the impact of stellar noise, in Table \ref{table:ratio1} we firstly compare it to the photon noise in terms of both $\sigma_{\tau}$ and $\sigma_{cr}$ for the two planets by finding the values at the timescale of $T_{14}$ for each.  The photometric stability requirement ('noise floor') for ARIEL is established at 100 ppm at the time scale of the planet transit, with a goal of 10 ppm (these are shown in Figures \ref{fig:comp1} and \ref{fig:comp2}). If a noise source falls below this requirement it should have negligible impact on the scientific objectives. The absolute noise on the contrast ratio due to stellar noise, $\sigma_{cr, sn}$, and the associated fractional uncertainty on the planet radius measurement, $\sigma_{R_p, sn}/R_p$, are given in Table \ref{table:ratio2}.  The latter is found through the error propagation formula,
 
\begin{equation}
\frac{\sigma_{R_p, sn}}{R_p} = \frac{1}{2}\frac{\sigma_{cr, sn}}{CR} \times 100\%
\end{equation}
\noindent where the contrast ratio, $CR = (R_p/R_s)^2$, and no error is assumed on $R_s$.

\vspace{-1.0em}
\section{Discussion}
The HD 209458 (G0V) star model gives consistently higher values for both $\sigma_{\tau,sn}$ and $\sigma_{cr,sn}$ than the GJ 1214 (M4.5V) model.  Both this and the higher fractional photon noise for the dimmer target, mean that the ratio of stellar pulsation and granulation noise to photon noise for GJ 1214b is much lower than for HD 209458b for both $\sigma_{\tau}$ and $\sigma_{cr}$ (Table \ref{table:ratio1}). In the mid-IR range (Ch0 and Ch1), the impact of pulsations and granulation is not significant when compared to ARIEL's photon noise for both GJ 1214b and HD 209458b, with the latter having the highest ratios, reaching 3.9\% for  $\sigma_{\tau}$ and 4.8\% for $\sigma_{cr}$ in Ch0.  This last result would increase the 1$\sigma$ error bar on the final spectrum of $(R_p/R_s)^2(\lambda)$ by only 0.1\% (compared to photon noise alone). The ratios are increased in the near-IR (NIRSpec) due to the combination of reduced fractional photon noise and increased fractional stellar pulsation and granulation noise, again much higher in HD 209458b where $\sigma_{cr,sn}$ is 25.1\% of $\sigma_{cr,pn}$. However this only increases the 1$\sigma$ error bar by just 3.1\%, so the impact remains small. In general, for the three visual range photometric channels still higher ratios occur both  for $\sigma_{\tau}$ and $\sigma_{cr}$.  However, even in these visual channels, the noise for GJ 1214b remains well below its photon noise (e.g. 4.6\% for $\sigma_{cr}$ in FGS 1).  By comparison, in HD 209458b, the stellar noise reaches a substantial proportion of, or is on par with, the photon noise (e.g. 105.8\% for $\sigma_{cr}$ in FGS 1, causing a 45.6\% increase in the 1$\sigma$ error bar), due mainly to the low fractional photon noise in these channels. The long $T_{14}$ for HD 209458b also means that photon noise integrates down longer, decreasing the relative difference with stellar noise which integrates down at a slower rate.  The low ratios seen in the VisPhot channel (0.5 \textmu m) for GJ 1214b, despite having the highest absolute stellar noise values for that system, can be explained by increased fractional photon noise as the signal falls in the Wien region of the M star SED.

\begin{table}
\begin{center}
\caption{Ratio of stellar pulsation and granulation noise to photon noise in $\sigma_{\tau}$ and $\sigma_{cr}$ at time scales of $T_{14}$ for GJ 1214b and HD 209458b.}
\label{table:ratio1}

\begin{tabular}{lccccc} 
\hline
\multicolumn{1}{c}{Channel} &
\multicolumn{2}{c}{GJ 1214b} &
\multicolumn{2}{c}{HD 209458b} \\

\multicolumn{1}{c}{(\textmu m)} &
\multicolumn{1}{c}{$ \frac{\sigma_{\tau,sn}}{\sigma_{\tau,pn}} $} &
\multicolumn{1}{c}{$\frac{\sigma_{cr, sn}}{\sigma_{cr, pn}}$} &
\multicolumn{1}{c}{$\frac{\sigma_{\tau, sn}}{\sigma_{\tau, pn}}$} &
\multicolumn{1}{c}{$\frac{\sigma_{cr, sn}}{\sigma_{cr, pn}}$} \\

\multicolumn{1}{c}{} &
\multicolumn{1}{c}{\%} &
\multicolumn{1}{c}{\%} &
\multicolumn{1}{c}{\%} &
\multicolumn{1}{c}{\%} \\

\hline

AIRS Ch1 (3.9-7.8) &0.4 & 0.4 & 3.7 & 4.5\\
AIRS Ch0 (1.95-3.9)&0.5 & 0.5 & 3.9 & 4.8\\
NIRSpec (1.25-1.9)&2.4 & 2.5 & 20.6 & 25.1\\
FGS 2 (1.1)&3.2 & 3.3 & 57.5 & 70.9\\
FGS 1 (0.9)&4.6 & 4.6 & 85.8 & 105.8\\
VisPhot (0.5)&1.1 & 1.2 & 88.2 & 104.8\\

\hline
\end{tabular}
\end{center}
\end{table}

\begin{table}
\begin{center}
\caption{Absolute $\sigma_{cr}$ due to stellar pulsation and granulation noise ($\sigma_{cr,sn}$) and associated fractional uncertainty on planet radius, 
$\sigma_{R_p, sn}/R_p$, for GJ 1214b and HD 209458b (assuming $R_p/R_s$ of 0.1162 for GJ 1214b and 0.1209 for HD 209458b).}
\label{table:ratio2}
\setlength{\tabcolsep}{5pt}
\begin{tabular}{lccccc} 
\hline
\multicolumn{1}{c}{Channel} &
\multicolumn{2}{c}{GJ 1214b} &
\multicolumn{2}{c}{HD 209458b} \\

\multicolumn{1}{c}{(\textmu m)} &
\multicolumn{1}{c}{$\sigma_{cr, sn}$} &
\multicolumn{1}{c}{$\frac{\sigma_{R_p, sn}}{R_p}$} &
\multicolumn{1}{c}{$\sigma_{cr, sn}$} &
\multicolumn{1}{c}{$\frac{\sigma_{R_p, sn}}{R_p}$} \\

\multicolumn{1}{c}{} &
\multicolumn{1}{c}{ppm} &
\multicolumn{1}{c}{\%} &
\multicolumn{1}{c}{ppm} &
\multicolumn{1}{c}{\%} \\

\hline
AIRS Ch1 (3.9-7.8) &	1.5 & 0.01 & 3.6 & 0.01 \\
AIRS Ch0 (1.95-3.9)& 2.3 & 0.01 &  3.5 & 0.01\\
NIRSpec (1.25-1.9)& 3.1 &  0.01  & 4.5 &  0.02\\
FGS 2 (1.1)&	 2.8  & 0.01 & 7.5 & 0.03 \\
FGS 1 (0.9)& 4.4 &  0.02 & 9.0 & 0.03\\
VisPhot (0.5)& 9.6 & 0.04 & 14.9 & 0.05\\

\hline
\end{tabular}
\end{center}
\end{table}

A strong atmospheric spectral feature on GJ 1214b would have a contrast ratio of $\sim$ 220 ppm (assuming a scale height of $\sim$ 30 km and a water-dominated atmosphere, and using the formula $A_p= 2[R_p5H]/{R_s}^2$, where $A_p$ is the atmospheric contrast ratio and $H$ is the scale height). For HD 209458b this will be $\sim$ 770 ppm (assuming a scale height of $\sim$ 510 km and a H$_2$-He atmosphere).  When comparing these values with the absolute uncertainties due to stellar pulsation and granulation noise in Table \ref{table:ratio2}, the impact on the transmission spectrum of each planet would be appear to be minimal at all wavelengths, with a maximum $\sigma_{cr,sn}$ of 14.9 ppm for HD 209458b and 9.6 ppm for GJ 1214b in the visual.  When considering $R_p$ in isolation the uncertainties due to stellar noise are very small (Table \ref{table:ratio2}), $\leq$ 0.05\% for both planets. 

\vspace{-0.8em}  
\section{Conclusions}

Noise from stellar variability arising from pulsation and granulation is unlikely to result in significant transit depth uncertainties in ARIEL's main spectroscopic channels (AIRS Ch0 and Ch1). Decorrelation should be unnecessary for most planets in these channels, Figure \ref{fig:comp1} showing a good separation between stellar pulsation and granulation noise and photon noise out to $\tau$ or $T_{14}$ of 10 hours.  Whilst the ratio of stellar noise to photon increases substantially in the visual range, for targets such as GJ 1214b, the stellar noise remains well below the photon noise.  The low impact is both due to increased fractional photon noise from a dim source, and decreased stellar granulation and pulsation amplitudes for the M-dwarf star.
For targets such as HD 209458b, the ratios are much higher, reflecting the low levels of fractional photon noise from the bright source, and higher amplitudes for both granulation and pulsation in the G-type star.  In addition, as the integration time increases with longer transits, such as for HD 209458b, the impact increases since the photon noise is beaten down at a faster rate than stellar noise. 
Noise from pulsations and granulation is unlikely to impact significantly on the transmission spectra of GJ 1214b or HD 209458b.  Stellar noise is well within the 'noise floor' requirement for ARIEL in all channels, and even below the 'goal' noise floor of 10 ppm at the timescales of the planet transits for all channels except VisPhot (with HD 209458b).   
Planets most likely to be impacted by stellar pulsation and granulation noise will be those with low fractional photon noise (e.g. around bright targets) with long transit times around Sun-like (rather than M-dwarf) stars, and that have small atmospheric scale heights, when measured in the visual range.   An example might be a terrestrial planet with a secondary atmosphere orbiting a near-by Sun-like star on a long period.  By changing the instrument model in ExoSim, this methodology can be easily applied to examine the stellar pulsation and granulation noise contributions to future JWST or E-ELT transmission spectroscopy observations.

\vspace{-0.8em}
\section*{Acknowledgements}

This work was supported by United Kingdom Space Agency (UKSA) grant ST/P002013/1. SS was funded through a Science and Technology Facilities Council (STFC) doctoral training grant.  IA and BV thank the European Space Agency (ESA) and the Belgian Federal Science Policy Office (BELSPO) for their support in the framework of the PRODEX Programme. The VIRGO instrument onboard SOHO is a cooperative effort of scientists, engineers, and technicians, to whom we are indebted. SOHO is a project of international collaboration between ESA and NASA.



\vspace{-0.5em}

\input{final2.bbl}
\bibliographystyle{mnras}

\bsp	
\label{lastpage}
\end{document}